\documentclass[useAMS,a4,usenatbib]{mn2e}
\input psfig.sty

\usepackage{floatflt, epsfig}

\def \xoff {\ifmmode x_{\rm off} \else $x_{\rm off}$ \fi}
\def \rhorms {\ifmmode \rho_{\rm rms} \else $\rho_{\rm rms}$ \fi}

\def \chisq  {\ifmmode  \chi^2   \else  $\chi^2$  \fi}  
\def \spose#1{\hbox  to 0pt{#1\hss}}  
\def \lta{\mathrel{\spose{\lower 3pt\hbox{$\sim$}}\raise  2.0pt\hbox{$<$}}}
\def \gta{\mathrel{\spose{\lower  3pt\hbox{$\sim$}}\raise 2.0pt\hbox{$>$}}}

\def \kms {\ifmmode  \,\rm km\,s^{-1} \else $\,\rm km\,s^{-1}  $ \fi }
\def \kpc {\ifmmode  {\rm~kpc}  \else ${\rm~kpc}$\fi}  
\def \pc {\ifmmode  {\rm~pc}  \else ${\rm~pc}$ \fi  }  
\def \Gyr {\ifmmode  {\rm~Gyr}  \else ${\rm~Gyr}$\fi}
\def \Msun {\ifmmode M_{\odot} \else $M_{\odot}$ \fi} 
\def \Lsun {\ifmmode L_{\odot} \else $L_{\odot}$ \fi} 
\def \Rsun {\ifmmode R_{\odot} \else $R_{\odot}$ \fi} 
\def \Msunpyr {\ifmmode M_{\odot}{\rm~yr}^{-1} \else $M_{\odot}{\rm~yr}^{-1}$ \fi} 
\def \hMsun {\ifmmode h^{-1}\,\rm M_{\odot} \else $h^{-1}\,\rm M_{\odot}$ \fi}

\def \LCDM {\ifmmode \Lambda{\rm CDM} \else $\Lambda{\rm CDM}$ \fi}
\def \sig8 {\ifmmode \sigma_8 \else $\sigma_8$ \fi} 
\def \OmegaM {\ifmmode \Omega_{\rm M} \else $\Omega_{\rm M}$ \fi} 
\def \OmegaL {\ifmmode \Omega_{\rm \Lambda} \else $\Omega_{\rm \Lambda}$\fi} 
\def \Deltavir {\ifmmode \Delta_{\rm vir} \else $\Delta_{\rm vir}$ \fi}
\def \rhocrit {\ifmmode \rho_{\rm crit} \else $\rho_{\rm crit}$ \fi}
\def \rhou {\ifmmode \rho_{\rm u} \else $\rho_{\rm u}$ \fi}
\def \zc {\ifmmode z_{\rm c} \else $z_{\rm c}$ \fi}

\def \rhos {\ifmmode \rho_{\rm s} \else $\rho_{\rm s}$ \fi} 
\def \rs {\ifmmode r_{\rm s} \else $r_{\rm s}$ \fi} 
\def \cvir {\ifmmode c_{\rm vir} \else $c_{\rm vir}$ \fi} 
\def \Rvir {\ifmmode r_{\rm vir} \else $R_{\rm vir}$ \fi}
\def \Vvir {\ifmmode V_{\rm  vir} \else  $V_{\rm vir}$  \fi} 
\def \Mvir {\ifmmode M_{\rm  vir} \else $M_{\rm  vir}$ \fi}  
\def \Nvir {\ifmmode N_{\rm  vir} \else $N_{\rm  vir}$ \fi}  
\def \Jvir {\ifmmode J_{\rm vir} \else $J_{\rm vir}$ \fi} 
\def \Evir {\ifmmode E_{\rm vir} \else $E_{\rm vir}$ \fi} 
\def \vvir {\ifmmode v_{\rm vir} \else $v_{\rm vir}$ \fi} 
\def \lam {\ifmmode \lambda  \else $\lambda$ \fi} 
\def \lamp {\ifmmode \lambda^{\prime} \else $\lambda^{\prime}$  \fi} 
\def \Vmax {\ifmmode V_{\rm  max} \else  $V_{\rm max}$  \fi} 

\def \Mgas {\ifmmode M_{\rm gas} \else $M_{\rm gas}$ \fi} 
\def \Mdisc {\ifmmode M_{\rm disc} \else $M_{\rm disc}$ \fi} 
\def \Md {\ifmmode M_{\rm d} \else $M_{\rm d}$ \fi} 
\def \Mda {\ifmmode M_{\rm d,0\%} \else $M_{\rm d,0\%}$ \fi} 
\def \Mdb {\ifmmode M_{\rm d,20\%} \else $M_{\rm d,20\%}$ \fi} 
\def \Mdc {\ifmmode M_{\rm d,40\%} \else $M_{\rm d,40\%}$ \fi} 
\def \md {\ifmmode m_{\rm d} \else $m_{\rm d}$ \fi} 
\def \Mb {\ifmmode M_{\rm b} \else $M_{\rm b}$ \fi} 
\def \Mbh {\ifmmode M_{\rm b,pri} \else $M_{\rm b,pri}$ \fi} 
\def \Mbs {\ifmmode M_{\rm b,sat} \else $M_{\rm b,sat}$ \fi} 
\def \zo {\ifmmode z_{0} \else $z_{0}$ \fi} 
\def \rd {\ifmmode r_{\rm d} \else $r_{\rm d}$ \fi} 
\def \rb {\ifmmode r_{\rm b} \else $r_{\rm b}$ \fi} 
\def \rbh {\ifmmode r_{\rm b,pri} \else $r_{\rm b,pri}$ \fi} 
\def \rbs {\ifmmode r_{\rm b,sat} \else $r_{\rm b,sat}$ \fi}

\title[Disc thickening with Gas] 
{Can Gas prevent the Destruction of Thin Stellar Discs by Minor Mergers?}

\author[B.P. Moster et al.] {Benjamin P. Moster$^{1}$
 \thanks{moster@mpia.de}, Andrea V. Macci\`o$^{1}$, Rachel S. Somerville$^{2,3}$,
 \newauthor{Peter H. Johansson$^{4}$ and Thorsten Naab$^{4}$}\\ 
  $^1$ Max-Planck-Institut f\"ur Astronomie, K\"onigstuhl 17, 69117 Heidelberg, Germany\\ 
  $^2$ Space Telescope Science Institute, Baltimore MD 21218\\
  $^3$ Department of Physics and Astronomy, Johns Hopkins University, Baltimore MD 21218\\
  $^4$ Universit\"ats-Sternwarte M\"unchen, Scheinerstr. 1, 81679 M\"unchen, Germany\\
}

\begin{document} 
              
\date{\today}
              
\pagerange{\pageref{firstpage}--\pageref{lastpage}}\pubyear{2009} 
 
\maketitle 

\label{firstpage}
             
\begin{abstract}
  
We study the effect of dissipational gas physics on the vertical
heating and thickening of disc galaxies during minor mergers. We
produce a suite of minor merger simulations for Milky Way-like
galaxies. This suite consists of collisionless simulations as well as
hydrodynamical runs including a gaseous component in the galactic
disc.  We find that in dissipationless simulations minor mergers cause
the scale height of the disc to increase by up to a factor of
approximately two. When the presence of gas in the disc is taken into
account this thickening is reduced by 25\% (50\%) for an initial disc
gas fraction of 20\% (40\%), leading to a final scale height \zo
between 0.6 and 0.7 kpc (for a sech$^2$ profile), regardless of
the initial scale height.  We argue that the presence of gas reduces
disc heating via two mechanisms: absorption of kinetic impact energy
by the gas and/or formation of a new thin stellar disc that can cause
heated stars to recontract towards the disc plane.  We show that in
our simulations most of the gas is consumed during the merger and thus
the regrowth of a new thin disc has a negligible impact on the \zo of
the post merger galaxy.  Final disc scale heights found in our
simulations are in good agreement with studies of the vertical
structure of spiral galaxies where the majority of the systems are
found to have scale heights of $0.4\kpc\lta\zo\lta0.8\kpc$. We also
found no tension between recent measurements of the scale height of
the Milky Way thin disc and results coming from our hydrodynamical
simulations.  Even if the Milky Way did experience a recent 1:10
merger it is possible to reproduce the observed thin disc scale
height, assuming that the disc contained at least 20\% gas (similar to
the gas fraction today) at the time of the merger.  We conclude that
the existence of a thin disc in the Milky Way and in external galaxies
is not in obvious conflict with the predictions of the CDM model.

\end{abstract}


\setcounter{footnote}{1}

\section{Introduction}
\label{sec:intro}

The cold dark matter (CDM) theory provides a successful framework for
understanding structure formation in our Universe. Within this paradigm,
dark matter first collapses in small haloes, which merge to form
progressively larger haloes over time
\citep[e.g.][]{white1978,davis1985}.  Major (near-equal mass) and
minor (unequal mass) mergers are a generic feature of structure
assembly in the CDM or hierarchical picture, and numerical simulations
have shown that many of these merging structures survive within the
virialized region of larger host haloes
\citep[e.g.][]{klypin1999,reed2005, diemand2008,springel2008}. These
merger events are now widely believed to be responsible for shaping
many galaxy properties.  Major mergers play an important role in
transforming disc-dominated spiral galaxies into spheroids \citep[e.g.][]
{hernquist1993,naab2003} and triggering episodes of
enhanced star formation (SF) and active galactic nuclei (AGN)
\citep{hernquist1989,barnes1992,mihos1994,barnes1996,mihos1996,
springel2005b,johansson2009}. Minor mergers may explain the origin
of thick discs \citep{abadi2003,brook2004,read2008,kazantzidis2008},
and the diffuse stellar halo around galaxies may be produced via tidal
destruction of merging satellites \citep[e.g.][]{bullock2005,
murante2004,bell2007}.

On the other hand, this large population of merging satellites has
raised the question of whether mergers are {\em too} common in the CDM
scenario. Some studies have questioned whether CDM models can produce
a large enough population of ``bulgeless'' discs (or systems with very
low bulge-to-total ratios, $B/T \lta 0.2$)
\citep[e.g.][]{graham2008,weinzirl2008} and whether thin, dynamically fragile
discs such as the one observed in the Milky Way can survive this
bombardment by incoming satellites \citep[e.g.][]{toth1992,quinn1993,
walker1996,wyse2001,bournaud2007}. Clearly, there are three main
aspects to settling this question: first, we must understand the
statistics of galaxy mass accretion and merger histories in a CDM
universe; second, we need to understand the physics of how galaxies
are transformed by mergers with various mass ratios, orbits, and other
parameters; and finally, we need accurate and unbiased statistics for
the observed populations of the relevant objects \citep[cf.][]{kormendy2008}.

The availability of large, high-resolution N-body simulations has made
progress possible on the first of these aspects. For example,
\citet{stewart2008} recently carried out a detailed study of
statistics of merger events in galaxy-sized dark matter haloes. They
confirmed previous results based on semi-analytic methods
\citep{purcell2007,zentner2007}, finding that the majority of the mass
delivery into a dark matter halo of mass $M_h$ is due to systems with
masses $M_{\rm sat}=(0.03-0.3)M_h$. They also found that a large
fraction (95\%) of Milky-Way sized haloes have accreted a satellite
with a virial mass comparable with the total mass of the Milky Way disc and
approximately 70\% have accreted an object with more than twice this
mass since $z\sim 1$.

There have been numerous studies based on collisionless
simulations that have tried to quantify the
effects of these minor mergers on the thickness and stability of
stellar discs \citep[e.g.][]{velazquez1999,font2001,benson2004,
gauthier2006,kazantzidis2008,kazantzidis2009,villalobos2008}.
This work has demonstrated that the answer depends quite sensitively
on the mass ratio of the satellite to the primary; there seems to be a
consensus that the main danger to thin discs is not from the
ubiquitous mergers with very small mass ratios (less than $\sim$ 1:10)
but rather from the rarer, yet still frequent, events with larger mass
ratio ($\gta$1:10). For example \citet{kazantzidis2008} studied the
effects of mergers with $(0.2\--1) M_{\rm disc}$ and concluded that
thin discs could survive such mergers. Employing dissipationless simulations,
\citet[][P09]{purcell2009} studied the response of a fully formed
Milky Way-like stellar disc within a $\sim 10^{12} \Msun$ DM halo to
mergers involving satellites with a total mass $M_{\rm sat} \sim
10^{11} \Msun \simeq 3 \, M_{\rm disc}$. They came to the conclusion
that regardless of the orbital configuration of the merger these
events transform the discs into structures that are roughly three times
thicker than the observed thin disc component of the Milky Way.

On the observational side, there is a general consensus that
$\sim$70\% of Milky-Way-mass ($ \sim 10^{12} \Msun$) haloes host a
disc dominated, late type galaxy \citep{weinmann2006,
  vdbosch2007,choi2007,park2007}. While it is also well known that the
majority of the mass in the disc of our Milky Way resides in a thin
component \citep[e.g.][]{juric2008}, how typical this situation is for
other disc galaxies in the Milky Way's mass range is less certain. Quantifying
the vertical thickness of the discs of external galaxies has been
attempted in a few studies \citep[e.g.][]{schwarzkopf2000,
  yoachim2006}, but is challenging because of small sample sizes,
inclination effects and extinction.

P09 suggested two possible explanations for the existence of a thin
disc in our Galaxy: one possibility is that the Milky Way is not a
representative case and that perhaps it has had an unusually quiet
accretion history for a halo of its mass. A second explanation is
related to the fact that all the numerical studies performed so far
have only considered the dissipationless components in the galaxy
(dark matter and stars), neglecting the presence of a dissipative gas
component in the disc. However, the inclusion of gas physics is known
to play an important role in stabilizing galactic discs.  Numerical
simulations have shown that gas is important for the survivability and
the regrowth of stellar discs during major mergers
\citep{barnes2002,springel2005c,robertson2006,naab2006,
  scannapieco2008,governato2009}, and that the presence of gas in
disc-type merger progenitors greatly suppresses the formation of a
post-merger spheroidal component \citet{hopkins2009a,hopkins2009b}. It
is reasonable to expect that the presence of gas in the progenitor
disc could also have an impact on the efficiency of the disc heating
and thickening: gas may be able to absorb some of the kinetic impact
energy of the merging satellite, and then radiate this energy away by
cooling; or, gas may be able to cool and reform a new thin disc after
the merger, forcing heated stars to contract again onto the disc
plane.

The main goal of this paper is to study in detail how the presence of
a dissipational gas component affects disc thickening in minor
mergers. For the first time, we address this problem using a suite of
high-resolution, fully hydrodynamical numerical merger simulations,
run with the Smoothed Particle Hydrodynamics (SPH) code {\sc GADGET-2}. We
consider simulations with and without gas, with a variety of initial
orbital parameters, and with different values of the star formation
efficiency parameters. We then re-examine the issue of whether the
cosmologically expected rate of minor mergers presents a problem for
CDM in light of our new results and the available observations.
 
The remainder of the paper is organized as follows: in
Section~\ref{sec:ncode} we provide a brief summary of the {\sc GADGET-2} code
and describe our simulations as well as the initial conditions for our
progenitor galaxies. In Section~\ref{sec:resu} we present our main
results, focusing on the difference between the dissipational and
dissipationless simulations as well as on the influence of different
initial conditions and star formation parameters on the final
thickness of the disc. Finally in Section~\ref{sec:conc} we summarize
and discuss our results, compare them with observations, and present
our conclusions regarding the issue of thin disc survival in a CDM
universe.

\begin{table*}
 \centering
 \begin{minipage}{140mm}
  \caption{Parameters kept constant for all simulations. Masses are in units of  $10^{10}\Msun$; scale and 
softening lengths are in units of\kpc~and\pc respectively.}
  \begin{tabular}{@{}lrrrrrrrrrrr@{}}
  \hline
  System & \Mvir & \Mdisc & \Mb & \rd & \rb & $c$ & $N_{\rm halo}$ & $N_{\rm disc}$ & $N_{\rm bulge}$ & $\epsilon_{\rm DM}$ & $\epsilon_{\rm disc}$\\
 \hline
 \hline
Primary & 100 & 2.4 & 0.600 & 3.0 & 0.5 & 9.65 & 4 000 000 & 1 000 000 & 500 000 & 100 & 50\\
Sat & 10 & 0.0 & 0.063 & 0.0 & 0.3 & 11.98 & 900 000 & 0 & 100 000 & 100 & 50\\
\hline
\label{t:conpar}
\end{tabular}
\end{minipage}
\end{table*}

\newpage

\section{Numerical Simulations} 
\label{sec:nsim}

\subsection{Numerical Code} 
\label{sec:ncode}

We make use of the parallel TreeSPH-code {\sc GADGET-2}
\citep{springel2005a} in this work. The code uses Smoothed Particle
Hydrodynamics \citep[SPH;][] {lucy1977,gingold1977,monaghan1992} to
evolve the gas using an entropy conserving scheme
\citep{springel2002}. Radiative cooling is implemented for a
primordial mixture of hydrogen and helium following \citet{katz1996}
and a spatially uniform time-independent local UV background
in the optically thin limit \citep{haardt1996} is included.

The SPH properties of the gas particles are averaged over the standard
{\sc GADGET-2} kernel using $\sim64$ SPH particles. Additionally the minimum
SPH smoothing length is required to be equal to the gravitational
softening length in order to prevent artificial stabilization of small
gas clumps at low resolution \citep{bate1997}. All simulations have
been performed with a high force accuracy of $\alpha_{\rm
  force}=0.005$ and a time integration accuracy of $\eta_{\rm
  acc}=0.02$ \citep[for further details see][]{springel2005a}.

Star formation and the associated heating by supernovae (SN) is
modelled following the sub-resolution multiphase ISM model described
in \citet{springel2003}.  The ISM in the model is treated as a
two-phase medium with cold clouds embedded in a hot component at
pressure equilibrium.  Cold clouds form stars in dense
($\rho>\rho_{\rm th}$) regions on a timescale chosen to match observations
\citep{kennicutt1998}.
The threshold density $\rho_{\rm th}$ is determined self-consistently by
demanding that the equation of state (EOS) is continuous at the onset
of star formation. We do not include SN-driven
galactic winds nor feedback from accreting black holes (AGN feedback)
in our simulations.

In our ``fiducial'' runs, we adopt the standard parameters for the
multiphase feedback model in order to match the Kennicutt Law as
specified in \citet{springel2003}.  The star formation timescale is
set to $t_*^0=2.1{\rm~Gyr}$, the cloud evaporation parameter to
$A_0=1000$ and the SN ``temperature'' to $T_{\rm SN}=10^8{\rm~K}$.  In
order to test whether the values of these (uncertain)
parameters affect our results, we reran two simulations with parameters
that are a factor of 4 larger ($t_*^0=8.4{\rm~Gyr}$, $A_0=4000$,
$T_{\rm SN}=4\times10^8{\rm~K}$). As \citet{springel2005b} noted, this
choice of parameters results in a star formation rate (SFR) of
$\sim 1~\Msun{\rm~yr}^{-1}$ for a Milky Way-like galaxy and gives
better agreement with the long gas consumption timescale inferred
for the Milky Way. However, these parameters yield a SFR that lies
slightly below the Kennicutt Law.  

\subsection{Galaxy Models} 
\label{sec:nmod}

\begin{table*}
 \centering
 \begin{minipage}{140mm}
  \caption{Parameters for the different simulation runs. Masses are in
    units of $10^{10}\Msun$, and scale height and softening lengths are in
    units of\kpc~and\pc respectively. The star formation timescale
    ($t_*^0$) is expressed in Gyr. The first three entries are for
    isolated galaxies, and the remainder are for mergers. }
  \begin{tabular}{@{}lrrrrrrrrrrr@{}}
  \hline
  Run & $f_{\rm gas}$ & \Mgas& \Md & $N_{\rm gas}$ & $\epsilon_{\rm gas}$ & \lam & \zo & $\theta$~ & $t_*^0$ & $A_0$ & $T_{\rm SN}$\\
 \hline
 \hline
IA & 0.0 & 0.0 & 2.4 & 0 & 0 & 0.033 & 0.40 & - & 2.1 & 1000 & $1\times10^8$K\\
IB & 0.2 & 0.6 & 3.0 & 125 000 & 70 & 0.034 & 0.40 & - & 2.1 & 1000 & $1\times10^8$K\\
IC & 0.4 & 1.6 & 4.0 & 333 333 & 70 & 0.036 & 0.40 & - & 2.1 & 1000 & $1\times10^8$K\\
\hline
MA60 & 0.0 & 0.0 & 2.4 & 0 & 0 & 0.033 & 0.40 & 60$^{\circ}$ & 2.1 & 1000 & $1\times10^8$K\\
MB60 & 0.2 & 0.6 & 3.0 & 125 000 & 70 & 0.034 & 0.40 & 60$^{\circ}$ & 2.1 & 1000 & $1\times10^8$K\\
MC60 & 0.4 & 1.6 & 4.0 & 333 333 & 70 & 0.036 & 0.40 & 60$^{\circ}$ & 2.1 & 1000 & $1\times10^8$K\\
\hline
MA45 & 0.0 & 0.0 & 2.4 & 0 & 0 & 0.033 & 0.40 & 45$^{\circ}$ & 2.1 & 1000 & $1\times10^8$K\\
MB45 & 0.2 & 0.6 & 3.0 & 125 000 & 70 & 0.034 & 0.40 & 45$^{\circ}$ & 2.1 & 1000 & $1\times10^8$K\\
\hline
MA30 & 0.0 & 0.0 & 2.4 & 0 & 0 & 0.033 & 0.40 & 30$^{\circ}$ & 2.1 & 1000 & $1\times10^8$K\\
MB30 & 0.2 & 0.6 & 3.0 & 125 000 & 70 & 0.034 & 0.40 & 30$^{\circ}$ & 2.1 & 1000 & $1\times10^8$K\\
 \hline
MA60T & 0.0 & 0.0 & 2.4 & 0 & 0 & 0.033 & 0.25 & 60$^{\circ}$ & 2.1 & 1000 & $1\times10^8$K\\
MB60T & 0.2 & 0.6 & 3.0 & 125 000 & 70 & 0.034 & 0.25 & 60$^{\circ}$ & 2.1 & 1000 & $1\times10^8$K\\
 \hline
MB60S & 0.2 & 0.6 & 3.0 & 125 000 & 70 & 0.034 & 0.40 & 60$^{\circ}$ & 8.4 & 4000 & $4\times10^8$K\\
MB60TS & 0.2 & 0.6 & 3.0 & 125 000 & 70 & 0.034 & 0.25 & 60$^{\circ}$ & 8.4 & 4000 & $4\times10^8$K\\
\hline
\label{t:varpar}
\end{tabular}
\end{minipage}
\end{table*}

We apply the method described in \citet{springel2005b} to construct
both the central galaxy (called the primary hereafter) and the merging
satellite systems.
Each primary system consists of gas and stellar discs with radial
profiles described by an exponential, with masses \Mgas and \Mdisc, and
a spherical bulge of mass \Mb embedded in a dark matter halo of mass
\Mvir. The halo has a \citet{hernquist1990} profile with a scale
radius $a$ corresponding to a Navarro-Frenk-White halo \citep[NFW;
][]{navarro1997} with a scale length of \rs and a concentration
parameter $c=\Rvir/\rs$.  We use the results of \citet{maccio2008} to
compute halo concentration as a function of virial mass.

The scale lengths \rd of the exponential gaseous and stellar discs are
assumed to be equal, and are determined using the model of
\citet{mo1998}, assuming that the fractional angular momentum of the
total disc $j_{\rm d}=(J_{\rm gas}+J_{\rm disc})/J_{\rm vir}$ is equal
to the global disc mass fraction $\md=(\Mgas+\Mdisc)/\Mvir$ for a
constant halo spin \lam. This is equivalent to assuming that the specific
angular momentum of the material that forms the disc is the same as
that of the initial dark matter halo, and is conserved during the process
of disc formation.

The vertical structure of the stellar disc is described by a radially
independent ${\rm sech}^2$ profile with a scale height $z_0$, and the
vertical velocity dispersion is set equal to the radial velocity
dispersion.  The vertical structure of the gaseous disc is computed
self-consistently as a function of the surface density by requiring a
balance of the galactic potential and the pressure given by the EOS.
The stellar bulge is constructed using the \citet{hernquist1990} profile with 
a scale length \rb.

Satellite systems are constructed using the same method adopted for
primary systems but they consist only of a dark matter halo and a
stellar bulge; neither stellar nor gaseous discs are included.

\subsection{Simulation Parameters} 
\label{sec:npar}

We construct a set of primary systems, each with a virial mass of
$\Mvir=10^{12}\Msun$ containing a disc and a bulge, and a
satellite system with a virial mass of $\Mvir=10^{11}\Msun$ containing
only a bulge. Since we wish to study a {\em typical} galaxy in a
$\Mvir=10^{12}\Msun$ halo, rather than use the specific parameters of
the Milky Way, we use the average stellar-to-halo mass ratio derived
by \citet{moster2009} to compute the stellar mass of every system
(both primaries and satellites). The \citet{moster2009} constraints
were derived empirically by asking how the population of dark matter
haloes and sub-haloes predicted by CDM must be populated with galaxies
in order to reproduce the observed galaxy stellar mass function, and
are in excellent agreement with constraints from other methods such as
galaxy clustering, satellite kinematics and weak lensing. Note that the
stellar mass we get for a typical galaxy in a $\Mvir=10^{12}\Msun$ halo
is significantly less than what is found for the Milky Way and M31
\citep[cf.][]{klypin2002}.

All primary systems have a concentration parameter of $c=9.65$ and a
stellar mass of $M_{\rm *,pri}=3 \times10^{10}\Msun$. Distributing
$80\%$ of this stellar mass into the exponential disc yields a stellar
disc mass of $\Mdisc=2.4\times10^{10}\Msun$ and a bulge mass of
$\Mbh=6\times10^{9}\Msun$.  We assume a bulge scale length of
$\rbh=0.5\kpc$.

\begin{figure*}
\psfig{figure=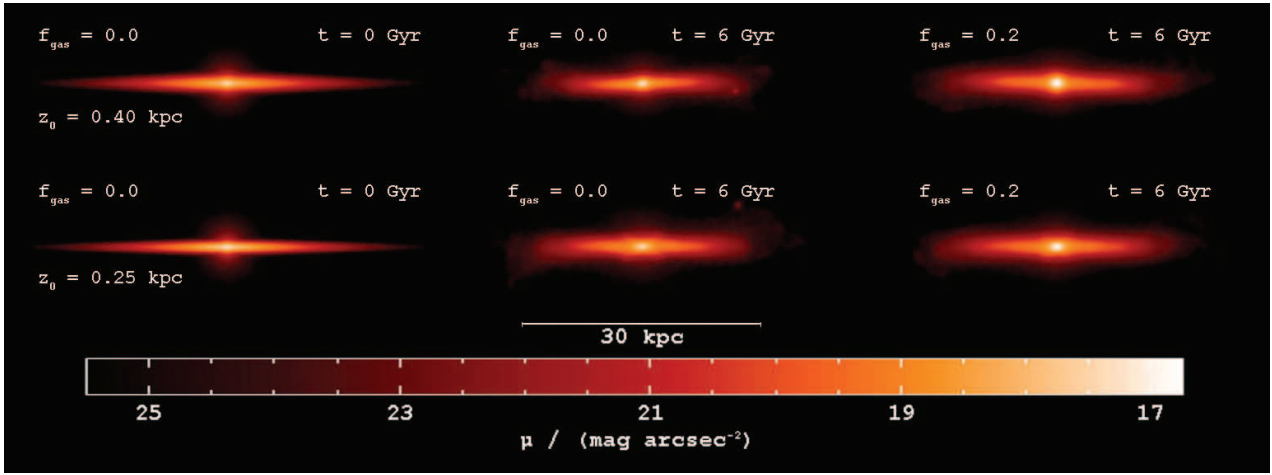,width=0.95\textwidth}
\caption{\scriptsize Edge-on surface brightness maps for galaxies with
  an initial scale height of $\zo=0.40\kpc$ (upper row) and
  $\zo=0.25\kpc$ (lower row). The left column shows the initial models
  while the centre and the right columns show the final galaxies
  (t=6\Gyr) for an initial gas fraction of 0\% and 20\%,
  respectively. A mass-to-light ratio of 3 $(M/L)_{\odot}$ has been
  assumed, typical of the Milky Way in the $B$-band.
}
\label{fig:maps}
\end{figure*}

For models that also include a gaseous disc component we add $M_{\rm
  gas, 20\%}=0.6\times10^{10}\Msun$ and $M_{\rm gas,
  40\%}=1.6\times10^{10}\Msun$ such that the gas fraction in the disc
is 20\% and 40\% in the two cases \citep[cf.][]{stewart2009}. 
The total disc mass is then
$\Mda=2.4\times10^{10}\Msun$, $\Mdb=3.0\times10^{10}\Msun$ and
$\Mdc=4.0\times10^{10}\Msun$ for a gas fraction of 0\%, 20\% and 40\%
respectively. We decided to keep the total stellar mass constant in order
to have a more direct comparison between the dissipational and
dissipationless cases. We fix the disc scale radius at $\rd=3.0\kpc$ for all
primary galaxies;
because in the \citet{mo1998} model, \rd depends on the disc mass and
on the halo spin parameter, we have choosen \lam in each case such
that this value of \rd is obtained in all primary systems.  The
corresponding spin parameters are $\lam_{0\%}=0.033$,
$\lam_{20\%}=0.034$ and $\lam_{40\%}=0.036$.  We consider two cases
for the initial disc scale height: we adopt a ``fiducial'' value of
$\zo=0.4\kpc$ for direct comparison with P09, and also
consider an initially thinner disc with $\zo=0.25\kpc$ for two values
of the gas fraction (0\% and 20\%).
Edge-on surface brightness maps for the initial conditions of our
primary galaxies are shown in the upper and lower left panels of
Figure \ref{fig:maps} for $\zo=0.4$ and $0.25\kpc$, respectively.

The satellite systems consist only of a dark matter halo and a
stellar bulge. The concentration parameter of the satellite halo is
$c=11.98$. Using the stellar-to-halo mass ratio we derive a stellar
mass of $\Mbs=6.3\times10^{8}\Msun$. We set the scale length of the
bulge to $\rbs=0.3\kpc$.

The primary systems always contain $N_{\rm dm}=4 \times 10^6$ dark
matter, $N_{\rm disc}=10^6$ stellar disc and $N_{\rm bulge}=5 \times
10^5$ bulge particles. Models with 20\% and 40\% gas fractions contain
$N_{\rm gas, 20\%}=1.25 \times 10^5$ and $N_{\rm gas, 40\%}=3.33
\times 10^5$ gas particles, respectively.  The satellite system
consists of $N_{\rm dm}=9 \times 10^5$ dark matter and $N_{\rm
  bulge}=10^5$ bulge particles. We set the gravitational softening
lengths to $\epsilon=50\pc$, $70\pc$ and $100\pc$ for stellar, gas and
dark matter particles, respectively.

Following P09 we choose orbits which are motivated by
studies of substructure accretion in cosmological N-body
simulations. The most likely values of the radial and tangential
velocity components ($v_r$ and $v_t$) are found to be respectively at
90\% and 60\% of the virial velocity of the primary halo
\citep{benson2005,khochfar2006}. For a halo of mass $\Mvir =
10^{12}\Msun$ the virial velocity is $\vvir=129\kms$ which results in
an initial subhalo velocity with $v_r=116\kms$ and $v_t=77\kms$. The
initial separation of the galaxies was chosen to be relatively large,
$d_{\rm start}=120\kpc$, in order to prevent significant perturbations
of the disc due to the sudden presence of the satellite's gravitational pull.
Once $d_{\rm start}$ is chosen, the orbital parameters are uniquely
determined. This, of course, limits our study to a single cosmologically
motivated orbit. Our combination of initial velocity and distance fixes the
pericentric distance at $\sim18\kpc$. We use a set of three orbital
inclinations ($\theta=60^{\circ},~45^{\circ} {\rm~and~} 30^{\circ}$, where
$\theta$ is the angle between the spin axes of the disc and the orbit) to
investigate the effect of the inclination on the rate of thickening of
the disc. All orbits are prograde and all simulations were evolved for
a total of $6{\rm~Gyr}$. Prograde orbits are expected to be more
destructive to the disc than retrograde mergers \citep{velazquez1999}.

We summarize the parameters that are kept constant for all simulations
in Table~\ref{t:conpar}, and parameters that differ for the various
simulation runs are summarized in Table \ref{t:varpar}. We label the
different simulations with the first letter I for isolated runs and an
M for mergers. The second letter signifies the gas fraction of the
disc followed by a number signifying the orbital inclination. We add a
T for discs with an initial scale height of $\zo=0.25$ and an S for
simulations with a lower star formation efficiency. We adopt a gas
fraction-inclination combination of 20\% and $60^{\circ}$ as our
fiducial model.

\section{Results}
\label{sec:resu}

\begin{figure}
\psfig{figure=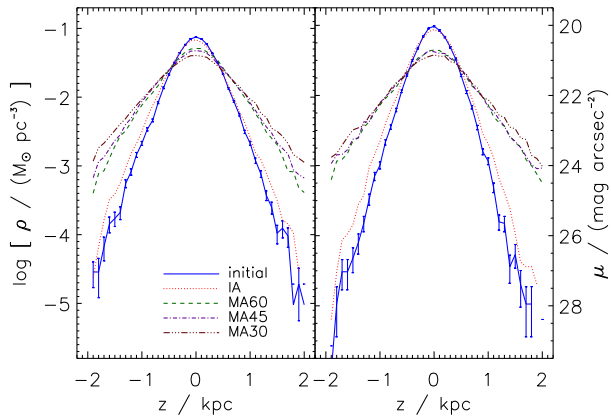,width=0.47\textwidth}
\caption{\scriptsize Density profiles for the initial ($z_0=0.4\kpc$) and
  final discs for an isolated galaxy and mergers with three
  inclination angles, for the simulations with no gas. The left panel
  shows the stellar mass density as a function of the distance to the
  galactic plane at a radius of $8\kpc$. The right panel shows the
  edge-on projected surface brightness at a projected radius of
  $8\kpc$.}
\label{fig:profiles}
\end{figure}

In order to study the evolution of the disc we compute the stellar
mass density as a function of the distance to the galactic plane at a
distance of $\Rsun\sim8\kpc$ from the disc spin
axis. Figure~\ref{fig:profiles} shows density profiles for a set of
simulations that do not contain any gas. In the left panel, stellar
mass density profiles of the initial disc (with $z_0=0.4\kpc$) and the
final discs for an isolated galaxy and mergers with three inclination
angles are plotted. To compare simulations to observations we also
compute the edge-on projected surface brightness as a function of the
distance to the galactic plane at a radius of $\Rsun$. This is done by
taking a vertical slice at a distance of \Rsun to the galactic centre
and assuming a mass-to-light ratio of $M/L=3(M/L)_{\odot}$,
appropriate for the Milky Way in the $B$-band \citep{zibetti2009}.
In the right panel of Figure~\ref{fig:profiles},
the edge-on surface brightness profiles are shown for the same
simulations. As is clearly visible from both panels, the disc of the
isolated galaxy is stable over the duration of the simulation, while
during mergers the profiles clearly broaden, implying that the disc
becomes thicker.

\begin{figure}
\psfig{figure=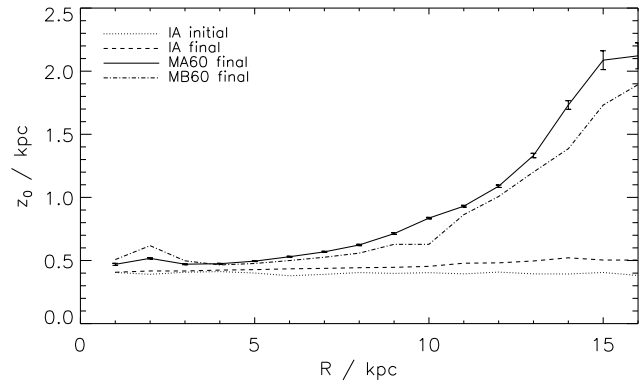,width=0.47\textwidth}
\caption{\scriptsize Disc scale height derived from fitting to the
  stellar mass density at different radii, for the initial (dotted
  line) and final states of a an isolated galaxy (IA; dashed line)
  and for a merger without gas (MA60, solid line) and with 20\%
  gas (MB60; dot-dashed line), both for an inclination of $60^{\circ}$. }
\label{fig:z0r}
\end{figure}

We note that the projected edge-on surface density profiles are
thicker than the ``slices'' through the 3d mass density profiles. To
understand this we plot the disc scale heights derived by fitting to
the 3d stellar mass density slices at different radii in Figure
\ref{fig:z0r}. This shows that \zo increases with increasing distance
from the galaxy center.  When computing the projected edge-on surface
density profiles, stellar particles with a larger distance from the
galactic centre than the specified radius are naturally included,
since some of these particles happen to lie along the line of
sight. These particles are on average also more distant from the
galactic plane, as Figure \ref{fig:z0r} indicates. This results in an
apparently thicker edge-on surface density profile and a larger
projected disc scale height. In order to compare our results with
observations we use the scale height obtained by fitting to the
projected edge-on surface density profile in the following.

\subsection{Stability of the Initial Conditions and Evolution of the scale height}
\label{sec:revo}

\begin{figure}
\psfig{figure=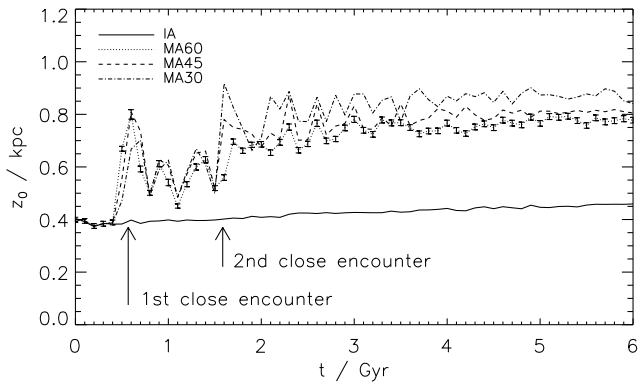,width=0.47\textwidth}
\caption{\scriptsize Evolution of the disc scale height for
  simulations with no gas. The solid line shows the scale height of an
  isolated galaxy while the other lines illustrate mergers with
  different inclinations.}
\label{fig:z0nogas}
\end{figure}

Galactic discs are fragile systems and there is always the possibility
that purely numerical effects can modify their morphology,
e.g. through bar formation or flaring. The stability of the initial
disc is then a key point to be addressed before looking at the effects
of satellite mergers.  To quantify the amount of disc thickening we
fit for the value of \zo assuming a ${\rm sech}^2$ profile. We always
fit to the projected edge-on surface brightness profile at \Rsun as
discussed above.

The dotted and dashed lines in Figure \ref{fig:z0r} show the values of the
disc scale height as a function of radius for an isolated galaxy (IA) at $t=0$
and $t=6\Gyr$, respectively. The final disc does not develop any appreciable
flaring and the thickening at larger distances is negligible compared to the
thickening due to satellite accretion events.
In Figure~\ref{fig:z0nogas} we show the resulting time evolution of \zo
for an isolated galaxy and mergers with varying inclinations, all
containing no gas. The isolated case (solid line) shows that in the
absence of perturbations the thin disc is extremely stable during the 6
Gyr timescale over which we run the simulations. This is also true for
the isolated runs including a gas component (IB and IC) as the solid and
dashed lines in Figure \ref{fig:z0wgasfrac} show. This implies that any
increase in \zo is due to accretion events. We see that regardless of
the inclination of the orbit, the disc thickens significantly in the
merger simulations, with \zo increasing by a factor of $\sim2$. 

We note that we also ran a simulation of an isolated galaxy using one
fourth the number of particles of our fiducial case (in all species)
and find that the scale height increases by 50\%. This demonstrates
the importance of numerical resolution in determining the disc
stability.

\subsection{The effect of gas in the disc}
\label{sec:rgas}

\begin{figure}
\psfig{figure=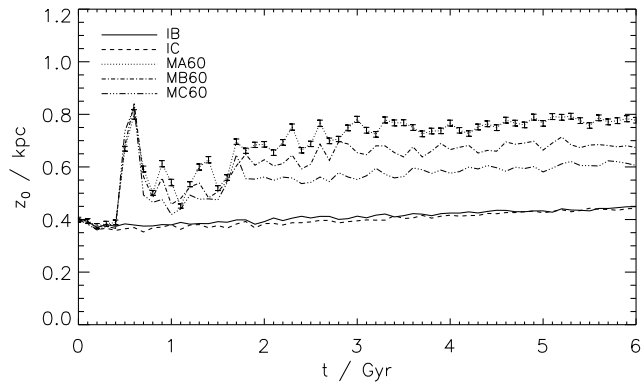,width=0.47\textwidth}
\caption{\scriptsize Evolution of the disc scale height for
  simulations with gas. The solid and the dashed lines show the isolated
   case while the other lines illustrate mergers with different gas fractions
   (0\%, 20\% and 40\% for MA60, MB60 and MC60, respectively).
}
\label{fig:z0wgasfrac}
\end{figure}

All of the previous numerical experiments devoted to studying the
heating and thickening of stellar discs by minor mergers have included
only the dissipationless components of the galaxies: dark matter and
stars. The presence of a gaseous component in the disc, which requires
a hydrodynamical approach to the problem instead of a purely
gravitational one, has been neglected so far.
The presence of gas may suppress disc thickening in two ways. One
possibility is that the gas may absorb some of the kinetic impact
energy of the merging satellite, which can then be removed from the
system by radiative cooling. In this way the impact energy that is
transferred to the stars, causing heating and thickening, may be
reduced. Another possible mechanism is that gas  forms a new thin
stellar disc after the merger. This disc could
then cause the heated stars to contract towards the disc plane.

\begin{figure}
\psfig{figure=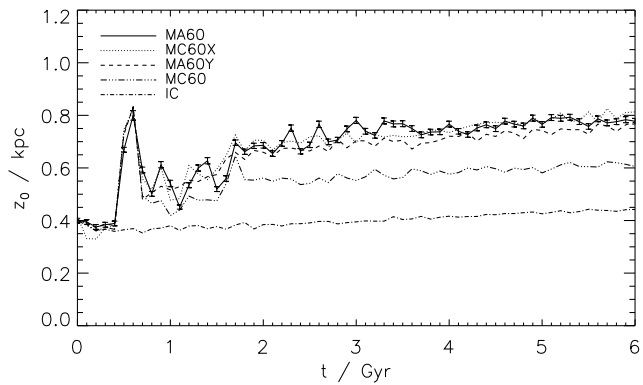,width=0.47\textwidth}
\caption{\scriptsize Evolution of the disc scale height for
  simulations without gas but with a more massive disc. The dotted
  lines shows the result for a simulation with an additional
  collisionless disc having the same mass and distribution as the gas
  disc in MC60. The dashed line shows the result of a simulation
  containing one stellar disc with increased mass (equal to the total
  baryonic mass in MC60). The solid and dot-dashed lines indicate the
  reference cases (MA60, MC60, IC). We see that the presence of
  dissipational gas, not the more massive disc, is responsible for the
  reduction in disc thickening. }
\label{fig:z0potential}
\end{figure}

We analyze the surface brightness profiles of the merger simulations
with gas, and show the resulting evolution of the scale height in
Figure~\ref{fig:z0wgasfrac} for initial gas fractions of 0\%, 20\% and
40\%, and an inclination of $60^{\circ}$ (MA60, MB60 and MC60). This
shows that the presence of gas does indeed suppress the thickening of
the disc by a minor merger. The final scale height increases by a
factor of $\sim 1.75$ for the 20\% gas case, and by only a factor of
1.5 for the 40\% gas case, in contrast to the factor of 2 increase in
the gas-free case (corresponding to a decrease in the final scale
height of 25\% and 50\%, respectively).

\begin{figure}
\psfig{figure=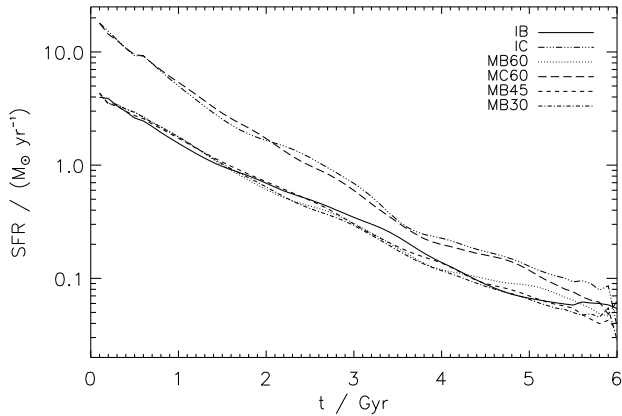,width=0.47\textwidth}
\caption{\scriptsize Star formation rates for isolated galaxies and
  mergers, with initial gas fractions of 20\% (IB, MB60, MB45, MB30)
  and 40\% (IC, MC60) in the progenitor disc. }
\label{fig:sfr1}
\end{figure}

One might wonder whether the reduced thickening is due to the
additional potential of the gas particles. In order to investigate
whether the gas physics or the greater potential have a larger
impact on the suppression of the disc thickening, we ran two
additional collisionless simulations. For the first simulation we
use the initial conditions of MC60 and convert all gas particles to
stellar particles at $t=0\Gyr$ (MC60X). We thus get an additional
stellar disc which has the same profile and potential as the
previous gaseous disc, but behaves collisionlessly during the
simulation. In a second simulation we create a galaxy like MA60 but
increase the mass of the stellar disc to
$\Mdisc=4.0\times10^{10}\Msun$ (MA60Y). This galaxy has the same
disc mass as MC60 and MC60X. Figure \ref{fig:z0potential} shows that
in both collisionless simulations (MC60X and MA60Y) the disc scale
height increases by a factor of $\sim2$. The final scale heights are
equal or even larger than for the less massive galaxy without gas
(MA60). This demonstrates that the suppression of the thickening in
the runs including gas is not due to the additional potential of the
gas disc but lies in the hydrodynamical nature of the particles.

In order to gain insights into the physical process that is causing
this change in behaviour, we examine the SFR as a function of
time. The results are shown in Figure~\ref{fig:sfr1}. The galaxy with
a gas fraction of 40\% (IC and MC60) naturally starts with a high SFR
($\sim18~\Msunpyr$), while the galaxy with a gas fraction of 20\% (IB,
MB60, MB45 and MB30) starts with a lower SFR of $\sim4~\Msunpyr$. We
see that 1:10 mergers do not have a noticable impact on the SFR.

The SFR quickly drops to low values under $1~\Msunpyr$ (after 2.5 Gyr
for a gas fraction of 40\% and 1.5 Gyr for the gas fraction of
20\%). This implies that most of the gas is consumed before the merger
is complete. Thus, the galaxy is not able to reform a new thin stellar
disc after the merger, which could pull heated stars towards the
galactic plane again. This means that the dominant process preventing
disc heating in the simulations above is the absorption of the kinetic
impact energy of the satellite by the gas component.
This can also be directly inferred from Figure~\ref{fig:z0wgasfrac}:
from about 2 Gyr, one can see that the growth of the scale height \zo
is slower in the simulations with gas, indicating that from the
beginning the gas is absorbing the impact energy and preventing the
disc from thickening considerably.
If the main mechanism was the reforming of a new thin disc, one would
see the disc thicken and then contract again. This would mean a
different slope for the \zo time evolution in runs with 
different gas fractions, however, we find similar slopes for all
gas fractions indicating that this process is not dominant.

\begin{figure}
\psfig{figure=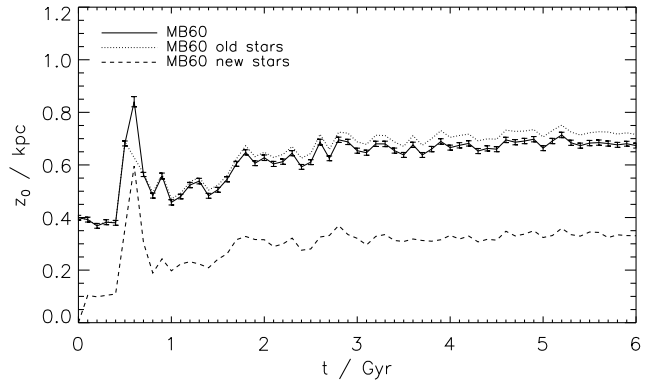,width=0.47\textwidth}
\caption{\scriptsize Evolution of the disc scale height for old stars
  (created in the initial conditions) and new stars (created during
  the simulation through star formation). The scale height for the
  combined sample is also shown for comparion.}
\label{fig:z0oldnew}
\end{figure}

\begin{figure}
\psfig{figure=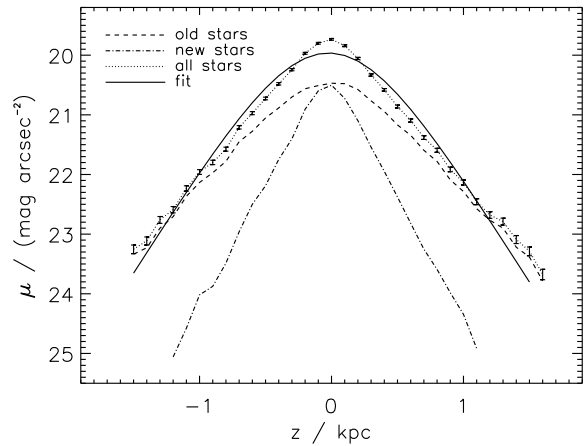,width=0.47\textwidth}
\caption{\scriptsize Final edge-on projected surface brightness profile
for the 40\% gas case at a projected radius of $8\kpc$ for old, new and
all stars. The solid line shows the fit to the combined stellar disc profile.}
\label{fig:profile40}
\end{figure}

Another way to demonstrate that the formation of a new disc does not
reduce the disc thickening noticeably in our simulations is to compare
the scale height of the stellar particles created in the initial
conditions (old stars) to the scale height of the stellar particles
that form during the simulation through SF (new stars). In
Figure~\ref{fig:z0oldnew} we compare the evolution of \zo for old and
new stars in our fiducial case MB60. The new stars clearly form a
thinner disc than the old stars, however, the combined sample has a
scale height which is only slightly thinner than the old stellar
disc. This is due to the much larger mass contained in the old stellar
disc component.
Figure~\ref{fig:profile40} shows the resulting edge-on projected
surface brightness profile for the 40\% gas case at $8\kpc$ for old, new
and all stars and a fit to the combined stellar disc using a ${\rm sech}^2$
model. The combined profile does not show an obvious transition
between the old and the new disc profile. However, one has to keep in mind
that we are using a single ${\rm sech}^2$ model to fit two distinct profiles,
implying that the fitted scale height of the combined sample decreases due to
the presence of the new thin disc. The new and old discs have
a final scale height of 0.38 and $0.66\kpc$ while the total stellar
disc has $\zo=0.61\kpc$. This shows that even for a gas fraction of
40\% (which is expected for disc galaxies at $z\sim1$ but not for local
galaxies at the present time), the new thin disc is not massive enough to
reduce the total disc scale height significantly.
In order to form a new thin disc with a mass comparable to the old
stellar disc, the SFR after the merger would have to be much higher.
This could occur in the case of an extremely gas-rich initial disc
\citep[$\geq 90$\% gas;][]{robertson2006,hopkins2008}.

\begin{figure}
\psfig{figure=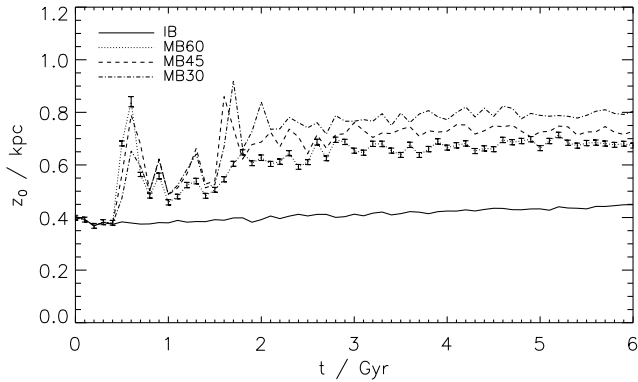,width=0.47\textwidth}
\caption{\scriptsize Evolution of the disc scale height for
  simulations with gas. The solid line shows the scale height of an
  isolated galaxy while the other lines illustrate mergers with
  different inclinations.}
\label{fig:z0wgasorbits}
\end{figure}

In order to investigate the influence of the orbital inclination on
the disc thickening we compute the evolution of the scale height for
mergers with gas fractions of 20\% for three different orbital inclinations
$\theta=60^{\circ},45^{\circ},30^{\circ}$(MB60, MB45, MB30
respectively). We plot the results in Figure~\ref{fig:z0wgasorbits}. 
The black solid line (IB) represents the isolated case with gas
(20\%), which is seen to be very stable as in the no-gas case.  The
results for different orbital inclinations show that the value of
$\theta$ has a small effect on the growth of the scale height.
Regardless of the inclination, the overall increase of the scale
height for simulations including gas is about 25\% smaller than in the
respective dissipationless simulations.

\subsection{A thinner initial disc}
\label{sec:rthi}

\begin{figure}
\psfig{figure=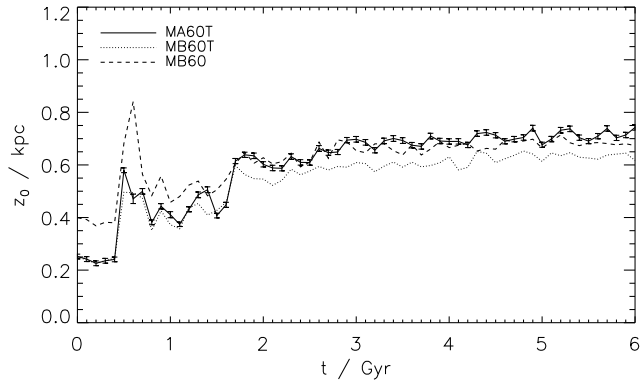,width=0.47\textwidth}
\caption{\scriptsize Evolution of the scale height for a disc with a
  smaller initial scale height. The solid line shows the `thin disc'
  simulation without gas, the dotted line shows thin disc case with
  20\% gas, and the dashed line shows the 20\% gas case with the
  original choice of initial scale height.}
\label{fig:z0thin}
\end{figure}

The choice of initial scale height for our primary disc was somewhat
arbitrary.  In order to investigate whether a thinner initial disc
would result in a thinner disc today, we simulate a merger of a disc
with an initial scale height of $\zo=0.25\kpc$ both without and with
gas (20\%) and an inclination of $60^{\circ}$. The resulting evolution
of \zo is shown in Figure~\ref{fig:z0thin}. The final scale height of
the disc without gas is as large as the scale height of the
corresponding initially thicker discs. This may be due to the fact
that an initially thicker disc is more robust to heating by accretion
events \citep{kazantzidis2009}. When gas is present, the final scale
height is slightly smaller in the case with an initially thinner disc,
but even a thinner initial disc is transformed into a system with a
scale height of more than 0.6\kpc. Thus, even when the effects of gas
in the progenitor disc are included, it seems to be difficult to
obtain a disc with a scale height that is substantially smaller than this
value following a minor (mass ratio greater than 1:10) merger.

\subsection{Lower SF efficiency}
\label{sec:rsfr}

In order to investigate the sensitivity of our results to the
parameters chosen for star formation and SN feedback, we have run some
simulations with lower star formation efficiency. In particular, we
were interested to see whether adopting a lower star formation
efficiency would allow a more significant new thin disc to reform
following the merger.  For this test we employ a star formation
timescale $t_*^0=8.4{\rm~Gyr}$, a cloud evaporation parameter
$A_0=4000$ and a SN ``temperature'' $T_{\rm SN}=4.0\times10^8{\rm~K}$
in order to have a SFR of $\sim 1~\Msun{\rm~yr}^{-1}$ for a Milky
Way-like Galaxy \citep{springel2005b}. We run this case for initial
scale heights of $\zo=0.4\kpc$ and $\zo=0.25\kpc$, in both cases using
our fiducial value of a 20\% gas fraction and an inclination of
$60^{\circ}$.

\begin{figure}
\psfig{figure=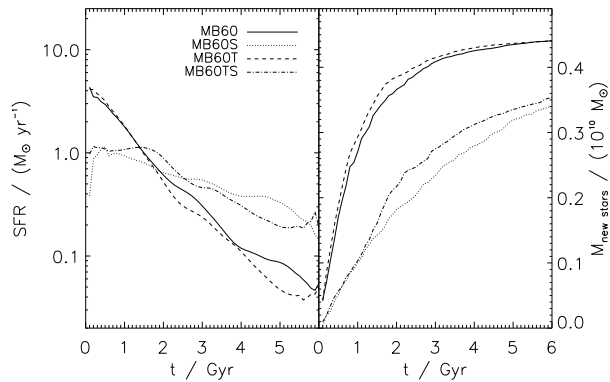,width=0.5\textwidth}
\caption{\scriptsize Left panel: SFR for simulations with the default
  (MB60, MB60T) and lowered SF efficiencies (MB60S, MB60TS), for the
  disc with initial scale height $\zo=0.4\kpc$ and $\zo=0.25\kpc$
  (T). Right panel: Mass of stellar particles formed during the
  simulations.}
\label{fig:sfr2}
\end{figure}

\begin{figure}
\psfig{figure=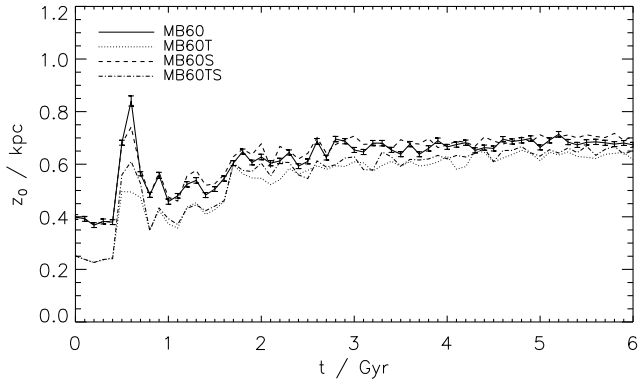,width=0.47\textwidth}
\caption{\scriptsize Evolution of the disc scale height for
  simulations with higher (MB60, MB60T) and lower SF efficiencies
  (MB60S, MB60TS).}
\label{fig:z0slow}
\end{figure}

We first compare the SFRs of the simulations with lowered SF
efficiencies with our fiducial results. The left panel of
Figure~\ref{fig:sfr2} shows the SFRs for the simulations MB60, MB60T,
MB60S and MB60TS. As we see, the SFRs in MB60S and MB60TS are
initially lower than those of MB60 and MB60T, respectively. After the
merger ($t\sim1.8{\rm~Gyr}$) however, the simulations with the lower
SF efficiency have a higher SFR, since MB60 and MB60T have already
consumed most of their gas supply. Due to the lower SFR during the
early stages of the simulation, the amount of newly formed stellar
mass is still lower in MB60S and MB60TS, as can be seen in the right
panel of Figure~\ref{fig:sfr2}. This means that the thin disc
consisting of new stars affects the overall disc scale height even
less than in the simulations MB60 and MB60T. The resulting scale
heights are plotted in Figure \ref{fig:z0slow}. As expected from the
SFRs, the scale heights of MB60S and MB60TS are slightly larger than
those of MB60 and MB60T, respectively.

Clearly, in order for the disc thickening problem to be solved by the
growth of a new stellar disc following the merger, the mass of the new
disc must be comparable in mass to that of the old stellar disc. This
implies that a significant amount of gas must remain available at the
end of the merger. Based on our results, when the SF and SN feedback
parameters are tuned be consistent with the empirical Kennicutt law,
most of the gas tends to be consumed during the merger. Simulations
with much higher initial gas fractions ($\sim 90$\%) or which included
a new source of cold gas (e.g. via cooling and accretion from a hot
halo) might be able to regrow a more massive thin disc.

\section{Conclusions and discussion}
\label{sec:conc}

We investigated the role of dissipational gas physics in the vertical
heating and thickening of disc galaxies by minor mergers (mass ratio
$\sim$ 1:10). We used the parallel TreeSPH-code {\sc GADGET-2} to simulate a
suite of minor merger simulations for a Milky Way-like primary
galaxy. The suite consists of collisionless simulations as well as
runs containing gas in the disc. Using cosmologically motivated
orbital parameters, we ran simulations with three different orbital
inclinations ($60^{\circ}, ~45^{\circ} {\rm~and~} 30^{\circ})$ and two
different initial disc scale heights ($\zo=0.4\kpc$ and
$\zo=0.25\kpc$).

We fit for the value of the scale height $z_0$, assuming a ${\rm
  sech}^2$ form for the vertical density profile. We showed that the
scale height derived from a fit to the mass density is a strong
function of the radial distance to the disc axis: as is well known,
minor mergers tend to cause flaring in the outer part of the
disc. This results in projected edge-on surface density profiles which
appear thicker than the actual three dimensional density profiles
since stars that are at a large distance from the galactic centre and
thus also more distant from the galactic plane are viewed along the
line of sight.

We found that in dissipationless simulations, minor mergers (without
gas) caused the scale height of the disc to increase by about a factor
of two, with the orbital inclination affecting the amount of thickening
by only about five percent. Thus, in qualitative agreement with the
results of P09, we find that in the absence of gas, cold thin discs
are destroyed by minor mergers. However, it is interesting that in
spite of the fact that the parameters of our simulations were chosen
to be very similar to those of P09, they found that the scale height
of the disc increased by a larger factor of $\sim 3$, leading to a
final disc about twice as thick as ours (despite having the same
initial scale height).

Although the discs in both simulation suites are stable against secular
evolution effects, this does not necessarily ensure that both models are
equally resistant to strong perturbative effects during an accretion event.
Possible reasons for this discrepancy may be the use of a different
IC generator and a different simulation code. The velocity dispersions
for example are fixed differently; while in our ICs the radial and the vertical
velocity dispersions are equal, the code used by P09 reproduces the
velocity ellipsoid for the Milky Way. This may result in an initially hotter 
disc for our typical galaxy (for a $\Mvir=10^{12}\Msun$ halo) which is more
resilient to massive accretion events.
Other intrinsic differences in the simulations of P09 are a more massive
disc (which may potentially be less stable due the formation of a bar)
and a larger stellar mass of the satellite (which results in a larger gravitational
pull each time the subhalo passes the disc).
We also note that P09 use a different concentration parameter (for both the
primary halo and the subhalo) and chose a Sersic profile for the bulge and an
NFW profile for the halo component while we use a Hernquist profile for both
components.
It is not obvious which of the mentioned differences has the largest impact
on the resulting disc scale heights.

We note that our results are consistent with those of
\citet{villalobos2008}.  These authors start with similar initial
conditions but employ a thicker disc (\zo = 0.7\kpc). The final scale
heights are between 1.0 and 1.3\kpc, depending on the orbital
inclination and are not as extreme as found by P09. The relative
thickening is even lower than in our simulations, confirming that
initially thicker discs are more robust to accretion events.

We also investigated mergers in which the progenitor disc initially
contained a gas fraction of 20 or 40 \%. We found that the presence of
gas reduces the final scale heights by 25\% (50\%) for a gas fraction
of 20\% (40\%).  The final scale heights were between 0.6 and
0.7\kpc~(1.5 to 1.75 times larger than the initial value), depending on
the initial gas fraction and the orbital inclination.  We argued that
the presence of gas can have an impact on disc thickening via two
different mechanisms. One process is the absorption of kinetic impact
energy by the gas. This energy can then be dissipated via
radiation. Another possible effect is the formation of a new thin
disc, which can cause heated stars to recontract towards the disc
plane. We showed that in our simulations, in which the SF and SN
feedback parameters were set to be consistent with the empirical
Kennicutt law for the initial disc, most of the gas is consumed during
the merger, and therefore the regrowth of new thin discs has a
negligible impact on the mass-weighted scale height of the post-merger
galaxy. Therefore, it seems that the main process that suppresses disc
thickening in the presence of gas in our simulations is the absorption
of impact energy by the gas. We intend to investigate the details of
the physics of this process in more detail in future work.

We computed scale heights for old stars (i.e. stellar particles that
were present in the initial conditions) and for new stars (stellar
particles created during the simulation through SF). The scale height
for old stars was found to be slightly higher than the overall
value ($\sim0.73\kpc$ vs $\sim 0.67$ for our fiducial case of 20\%
gas and a $60^{\circ}$ inclination) while the scale height of the new
stars was about half of that value ($\sim0.35\kpc$). We showed that
the final mass of the new disc is small compared to that of the old
disc. This indicates that although a new thin disc does form, it does
not have a significant effect on the final thickness of the total
disc. We argue that in order to have a noticeable effect, the new thin
disc would need to be comparable in mass to the old disc, which would
require that a significant mass of cold gas is still present at the
end of the merger.

We ran simulations with two different initial scale heights,
$\zo=0.40\kpc$ (fiducial case) and $\zo=0.25\kpc$ (thin). As it turns
out, thinner discs are more unstable to heating and are therefore
thickened more by the merger than initially thicker discs. As a
result, the two cases result in discs with nearly the same final scale
height ($\sim 0.6\kpc$ for the 20\% gas fiducial case).

To study the sensitivity of our results to the parameters controlling
SF and SN feedback, we ran simulations with a lower star formation
efficiency. In these simulations, less gas is consumed and so the SFR
at the end of the merger is higher. However, we found that the final
mass of the new disc was still lower than in the simulations with the
higher SF efficiency. This results in final scale heights that are
slightly larger than in the fiducial case.  We conclude that in order
to reform a new thin disc comparable in mass to the old disc, either
the initial gas fraction would have to be much higher, or an external
fueling reservoir (such as cooling and accretion from a hot halo)
would be needed \citep{sommer-larsen2003, kaufmann2006,
peek2009, grcevich2009}

In light of our results, we can now reassess whether disc thickening
by minor mergers presents a serious problem for CDM. First, we need to
compare the fraction of disc galaxies that are expected to have had a
minor merger (mass ratio greater than 1:10 but less than about 1:4)
since $z\sim 1$ with the fraction of observed galaxies with thin
discs.  Observations show that roughly 70\% of Milky Way-sized haloes
host late-type galaxies \citep{weinmann2006,vdbosch2007}. Based on an
analysis of cosmological simulations, \citet{stewart2008} find that
20\% of Galaxy-sized haloes did not experience any merger event of 1:10
or larger in the past 8 Gyr. This implies that ~30\% of all Milky
Way-like galaxies did not have such a merger, and as a result can be
expected to have retained their thin disc. This means that if less
than 30\% of all Milky Way-like galaxies are found to have scale
heights that are substantially lower than those found in our
simulations ($\sim 0.6$ \kpc) then there is no discrepancy.

Obtaining statistically unbiased observational measurements of scale
heights for a complete sample of galaxies is difficult, due to small
sample sizes, dust extinction, and inclination effects. Still there
have been many studies focusing on the vertical structure of galaxies
\citep[e.g.][]{shaw1989,
  shaw1990,grijs1996,grijs1997,pohlen2000,kregel2002,bizyaev2004}. All
these studies suggest that the majority of observed galaxies with
properties similar to those of the Milky Way have ${\rm sech}^2$
scale heights in the range of $0.6<\zo<1.0\kpc$ with a ratio between
scale height and scale length between 0.2 and 0.3.
These values have been determined from observed two-dimensional
surface brightness profiles. Usually it is assumed that scale length and scale
height are independent parameters. A model 2d surface brightness profile
consisting of an exponential radial factor and a vertical factor
(${\rm sech}^2$, ${\rm sech}$ or ${\rm exp}$) is then fitted to the observed
2d profile. All scale heights cited here have been converted to a ${\rm sech}^2$
profile.

A statistical study of the vertical structure of spiral galaxies is
presented in \citet{schwarzkopf2000}. They found that the majority
($\sim60\%$) of galaxies have vertical scale heights less than
1.1\kpc, with a maximum between $0.4\kpc\leq\zo\leq0.8\kpc$, while
galaxies with a very thin disc ($\zo<0.4\kpc$) are extremely rare.
Similar results were obtained by \citet{yoachim2006}, who derived
scale heights for a sample of edge-on galaxies and presented a
relation between scale height and circular velocity. They showed that
Galaxy-sized systems are expected to have scale heights in the range
$0.6\kpc\leq\zo\leq1.2\kpc$.
In order to compare our results to these studies, we compute 2d edge-on
disc surface brightness profiles (as shown in Figure \ref{fig:maps}) and
fit a model 2d profile consisting of an exponential radial and ${\rm sech}^2$
vertical disc to the simulated profiles. For the collisionless fiducial run MA60
the final scale height is $0.74\kpc$, while for the fiducial runs including a
gas disc MB60 and MC60 the scale heights are $0.66$ and $0.62\kpc$,
respectively. These values fall precisely in the observed range.

Previous theoretical studies, such as P09, have focussed on whether
the observed scale height of the Milky Way is in conflict with the
predictions of CDM models. We revisit this issue as well. 
There have been various studies of the vertical structure of the Milky
Way \citep[e.g.][]{bahcall1980,kent1991, reid1993,larsen2003,juric2008}.
These studies find exponential scale heights for the Milky Way's old
thin disc of $h_z\sim0.3\kpc$ and for the young star forming disc of
$h_z\sim0.1\kpc$. As is standard in the literature, we convert these
exponential scale heights to values corresponding to a ${\rm sech}^2$
function using $\zo=2h_z$ (which yields very good agreement between
the two profiles at $z\geq\zo$ and reasonable agreement at
$\zo>z$). This results in scale heights of $\zo=0.6\kpc$ (old disc)
and $\zo=0.2\kpc$ (young disc). The former is just what we find for
the overall scale height in our fiducial merger with gas, while the
latter is very close to what we find for the disc of new stars, as
shown in Figure~\ref{fig:z0oldnew}. Thus, in light of our results,
even if the Milky Way did experience a recent 1:10 merger, if the disc
contained even 20\% gas (similar to the gas fraction today) at the
time of the merger, there is no conflict with observations. Of course,
it is also possible to explain the Milky Way by assuming that it is
one of the 30\% of systems that did not experience a 1:10 merger. It
has been previously suggested by \citet{hammer2007}, based on the
unusually low mass and angular momentum of the Milky Way disc, that
our Galaxy may have had a particularly quiescent formation history.

We also note that while we are fitting the scale height to mass
profiles, observational studies use luminosity profiles to derive
\zo. Based on the models of \citet{bruzual2003} for a single age
stellar population, we expect the $B$-band mass-to-light ratio of very
young stars ($t_a<2\Gyr$) and old stars ($t_a>8\Gyr$) to differ by a
factor of $\sim4$.  This implies that the new stellar particles should
carry a correspondingly larger weight in a luminosity-weighted fit,
resulting in a smaller estimated scale height. However, as seen in
Figure~\ref{fig:sfr2}, the mass of such young stars formed in our
simulations is very small (only about a tenth of the mass of the
pre-existing stellar disc), so even if we accounted for the
age-dependent mass-to-light ratio, the overall measured scale height
would not differ significantly from the mass-weighted one. Moreover,
young stars will also tend to be more enshrouded by dust, further
reducing their impact on the measured scale height.

Finally, the simulations presented here still neglect the larger scale
environment and cosmological growth of the galaxy. The progenitors of
Milky Way discs presumably had somewhat different properties (smaller
masses, higher gas fractions, and possibly thicker discs) than their
present-day counterparts. Perhaps most importantly, one expects these
systems to have accreted a significant amount of new material over the
course of the past 6--8 Gyr via accretion and cooling from a hot gas
halo. This additional supply of cold gas to the disc could enhance the
formation of a new, thin disc and further reduce the thickening by
minor mergers. We plan to re-examine this problem in the context of
simulations with a cosmologically motivated merger and accretion
history, and including cooling from a hot halo, in a future work.

In final summary, in contrast with the conclusions reached by P09
using dissipationless simulations, we
conclude that the existence of a thin disc in the Milky Way is not in
obvious conflict with the predictions of the CDM model. Although the
Milky Way may have had an unusually quiescent merger history, mooting
the issue of disc thickening by minor mergers, our results indicate
that one does not need to rely on such an argument. When we include a
moderate amount of cold gas in the progenitor disc (similar to the gas
fractions observed in typical spirals today), the scale heights of
simulated discs that have experienced a 1:10 merger are in good
agreement with the observed values both for the Milky Way thin disc
and for external galaxies.

\section*{Acknowledgements} 

We thank Chris Purcell, Stelios Kazantzidis and James Bullock for
helpful comments on a draft version of this paper and for the
following fruitful discussion on the differences of our results and
their possible reasons.
We also thank Volker Springel for providing the code used to create
the initial conditions.
In addition we thank
Frank van den Bosch,
Jerry Ostriker
and Justin Read
for enlightening discussions and useful comments on this work.
The numerical simulations used in this work were performed
on the PIA cluster of the Max-Planck-Institut f\"ur Astronomie and
on the PanStarrs2 clusters at the Rechenzentrum in Garching. BPM thanks the
Space Telescope Science Institute for hospitality and financial
support for his visit. BPM also acknowledges a travel grant from the
German Research Foundation (DFG) within the framework of the
excellence initiative through the Heidelberg Graduate School of
Fundamental Physics.


\bibliographystyle{mn2e}
\bibliography{moster2009}

\label{lastpage}

\end{document}